\documentclass{svjour3}                     
\smartqed  
\usepackage{graphicx}
\journalname{Journal of Low Temperature Physics - QFS2009}

\begin{document}

\title{Studying the normal-fluid flow in Helium-II using metastable helium
molecules
}


\author{W. Guo \and J.D. Wright \and S.B. Cahn \and J.A. Nikkel \and D.N. McKinsey
}


\institute{W. Guo \and J.D. Wright \and S.B. Cahn \and J.A. Nikkel \and D.N. McKinsey
\at Physics Department, Yale University,\\ New Haven, CT 06520, USA\\
Tel.: 203-432-3825\\ Fax: 203-432-9710\\
\email{wei.guo@yale.edu} }

\date{Received: date / Accepted: date}

\maketitle

\begin{abstract}
We demonstrate that metastable helium molecules can be used as tracers to visualize the flow of the
normal fluid in superfluid $^{4}$He using a laser-induced-fluorescence technique. The flow pattern of a normal-fluid jet impinging on the center of a copper disc is imaged. A ring-shaped circulation structure of the normal fluid is observed as the jet passes across the disc surface. The fluorescence signal for the molecules trapped in the circulation structure is measured as a function of time after we turn off the molecule source. The radiative lifetime and density of the molecules can be determined by fitting the measured data using a simple analytic model. We also discuss a proposed experiment on using a previously developed molecule tagging-imaging technique to visualize the normal-fluid velocity profile during the transition of quantum turbulence in a thermal counterflow channel.
\end{abstract}

\section{Introduction}
Recently, particle image velocimetry with polymer micro-spheres and hydrogen isotopes has been used
to study liquid helium flows~\cite{Sciver JLTP 2005,Sciver Nat phys} and solid hydrogen tracers
have been used to visualize the quantized vortices~\cite{Bewley,Paoletti PRL}. However, the
dynamics of micron-sized tracers in the presence of vortices are complex~\cite{Sciver JLTP 2005}.
One must account for particle-vortex interactions~\cite{Kivotides2008-78} in order to
extract an accurate measurement of the local normal-fluid velocity. Another approach to image the flow in liquid helium is neutron absorption
tomography~\cite{Hayden2004}, which uses $^{3}$He as a neutral tracer and requires a
finely collimated neutron beam and the ability to raster the neutron beam through the region of
interest. Here we shall demonstrate that metastable He$_{2}^{*}$ triplet molecules as neutral tracers can be a powerful tool for helium hydrodynamic research. Metastable He$_{2}^{*}$ molecules can be imaged using a
laser-induced-fluorescence technique which involves only table-top laser
systems~\cite{McKinsey,Rellergert}. He$_{2}^{*}$ molecules follow the motion of the normal fluid
without being affected by vortices at temperatures above 1~K~\cite{Vinen} due to their small
effective mass in liquid $^{4}$He~\cite{Benderskii}, hence they are ideal tracers for the normal fluid. In the following demonstration experiment, the real time dynamics of a normal-fluid jet impinging on the center of a copper disc is studied by using the molecule visualization technique.
\section{Experiment and results}
\begin{figure}
\begin{center}
\includegraphics[scale=0.4]{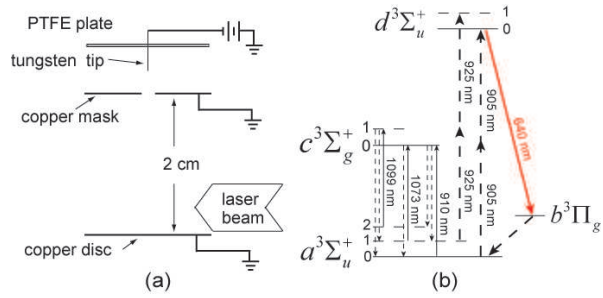}
\caption{(Color online) (a)~Schematic diagram showing the experimental setup. (b)~Schematic diagram showing the cycling transitions for imaging the
He$_{2}^{*}$ triplet molecules.}\label{laser}
\end{center}
\end{figure}
The schematic diagram of the experimental setup is
shown in Fig.~\ref{laser}~(a). A sharp tungsten tip was mounted at the center of a
polytetrafluoroethene (PTFE) plate. A grounded copper disc 2~cm in diameter and 0.5~mm thick was placed about 3~mm away from the tip apex as a mask. A hole 1~mm in diameter was cut at the center of the copper mask. A second grounded copper disk with no hole cut at its center was placed 2~cm below the copper mask. The whole device was held at the center of a 250~cm$^{3}$ helium cell and was thermally linked to a helium bath. The temperature of the system was controlled to be at about 2~K by directly pumping on the helium bath.

To produce He$_{2}^{*}$ molecules in the liquid helium, a negative voltage of amplitude higher than
the field-emission threshold is applied to the tip \cite{McClintock,Dahm}. Electrons emitted by the tungsten tip move from the tip region to the copper mask plate under the applied electric field. The moving electrons continuously pull the normal fluid and lead to the formation of a normal-fluid jet~\cite{Dahm}. Associated with the emitted electrons, He$_{2}^{*}$ molecules in both spin singlet and triplet states are produced near the tip apex. The singlet molecules radiatively decay in a few nanoseconds~\cite{Chabalowski}, while the triplet molecules are metastable with a radiative lifetime of about 13~s in liquid $^{4}$He~\cite{McKinsey1999}. At 2.0~K, a He$_{2}^{*}$
molecule diffuses less than 1~mm during its lifetime~\cite{McKinsey}. The generated metastable He$_{2}^{*}$ molecules are essentially entrained in the normal-fluid jet. As the jet reaches the surface of the mask plate, part of it can pass through the hole and keep on flowing to the bottom copper disc. Since both the mask and bottom disc were grounded in the experiment, few electrons could leak through the small hole in the mask. Hence, in the region between the mask and the bottom disc, the normal fluid moves without any electron pulling force on it, and its genuine hydrodynamic motion can be visualized by imaging the metastable He$_{2}^{*}$ molecules entrained.

To image the He$_{2}^{*}$ molecules in the triplet state, a single pulsed laser at 905~nm is used to drive the
molecules out of the $a^{3}\Sigma_{u}^{+}$ state to produce 640~nm fluorescent light through a cycling
transition (see Fig.~\ref{laser}~(b))~\cite{Rellergert}. However, during the cycling transition, He$_{2}^{*}$ molecules may fall to the long-lived excited vibrational levels of the $a^{3}\Sigma_{u}^{+}$ state and are lost for subsequent cycles~\cite{Rellergert,Rellergert thesis}. To recover the lost molecules, we use continuous fiber lasers at
1073~nm and 1099~nm to repump the molecules from the $a(1)$ to the $c(0)$ state and from
the $a(2)$ to the $c(1)$ state, respectively. Molecules in the $c$ states have a chance to decay
back to the $a(0)$ state and can be used again. In the experiment, the intensities of the fiber lasers at 1073~nm and 1099~nm were chosen to
be 3~W/cm$^{2}$ and 1.5~W/cm$^{2}$, respectively. The intensity of the pulsed laser at 905~nm was
500~$\mu$J/cm$^{2}$ per pulse, and the repetition rate was 500~Hz. The pulsed laser beam was overlapped with the continuous
laser beams in the region near the bottom disc and both were expanded to a spot size of 1~cm$^{2}$. All the laser beams pass through a pair of anti-reflection coated windows on the cell with negligible heating inside the cell.

\begin{figure}
\begin{center}
\includegraphics[scale=0.42]{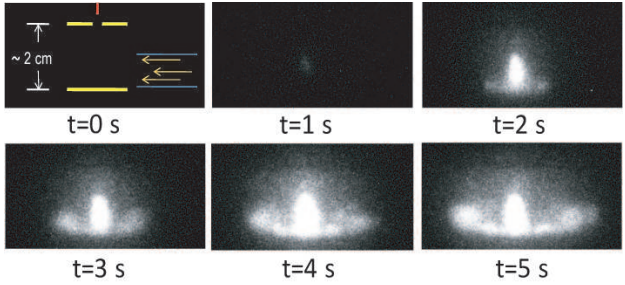}
\caption{(Color online) The motion of a normal-fluid jet impinging on the center of a copper disc. The images were taken at 0 to 5~s after a -800~V voltage was applied to the tungsten tip.}\label{jet image}
\end{center}
\end{figure}
A typical set of images showing the motion of a normal impinging normal-fluid jet on the center of the bottom disc is shown in Fig.~\ref{jet image}. These images were taken with an intensified CCD camera successively from 0~s to 5~s at one second time intervals after a -800~V voltage was applied to the tip. The camera was synchronized to each laser pulse and exposed for 6~$\mu$s so as to minimize dark current. The number of camera exposures for each image was chosen to be 50. As one can see, the normal-fluid jet enters the laser illumination region at t=1~s and a weak signal due to the jet front can be seen. The jet then impinges on the bottom disc with a velocity of about 1~cm/s and leads to a radial wall jet~\cite{M.B. Glauert} on the disc surface. As the radial wall jet passes the edge of the bottom disc, a ring-shaped circulation structure of the normal fluid appears near the periphery of the bottom disc. In our experiment, only the part of the vortex ring that was illuminated by the laser beams was visible. Hence the bright round regions near the edge of the bottom disc in the later images represent the cross-section view of the vortex ring. The impingement of a free circular jet onto a flat surface at normal incidence is of interest in many practical problems, such as jet blast, airplane vertical takeoff and landing, and particularly atmospheric microbursts~\cite{Fujita} which is an event where a mass of cool dense air falls to the ground and spreads horizontally and radially from the impingement site, often creating hurricane-force winds. In our experiment, the impinging velocity of the jet can be controlled easily by varying the electric current. We even developed a technique for measuring the local flow velocity along the jet by tagging a small group of He$_{2}^{*}$ molecules using a focused 910 nm pump laser pulse and imaging only the tagged molecules at a delayed time using an expanded 925 nm probe laser pulse. The pump laser drives the molecules from the $a(0)$ state to the $c(0)$ state where part of the molecules decay to the $a(1)$ state (see Fig.~\ref{laser}~(b)). The probe laser then drives the $a(1)$ molecules to the $d$ state to produce the fluorescent light~\cite{WeiPRL}. Our simple but convenient setup combined with the molecule detection technique may provide a new way to simulate and understand phenomena in a jet impinging process. A detailed study on jet impingement may be conducted in the future.

\begin{figure}
\begin{center}
\includegraphics[scale=0.48]{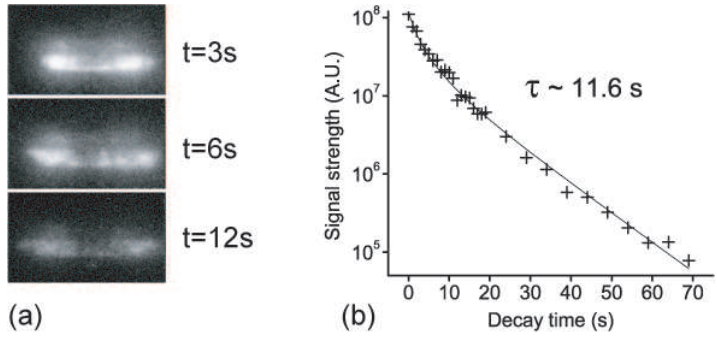}
\caption{(a)~Fluorescence images showing the He$_{2}^{*}$ molecules trapped in the normal-fluid circulation structure at 3~s, 6~s and 12~s after the tip was turned off. (b)~The collected fluorescence light in arbitrary unit as a function of time after the tip was turned off. The crosses represent the measured data. The solid line is a least-squares fit to the measured data using the formula as discussed in the text.}\label{molecule decay}
\end{center}
\end{figure}
As we turn off the voltage on the tungsten tip, the source of the normal-fluid jet is off immediately, but the ring-shaped circulation on the periphery of the bottom disc can last for a few more seconds. Typical images taken with the same 905~nm laser settings but at 3~s, 6~s and 12~s after the tip was turned off are shown in Fig.~\ref{molecule decay}~(a). The central bright feature with respect to Fig.~\ref{jet image} is absent since the jet is turned off. The He$_{2}^{*}$ molecules initially trapped in the core region of the circulation structure can stay there for long time even after the circulation dies away since there was no net normal-fluid flow to disperse the molecules. The fluorescence signal decreases gradually as the molecules decay with time. This indeed provides us a convenient way to study the decay process of the He$_{2}^{*}$ molecules. The crosses in Fig.~\ref{molecule decay}~(b) represent the measured total fluorescence light in arbitrary unit as a function of the decay time. The decay process of the He$_{2}^{*}$ molecules in the experiment can be simply described by the following equation
\begin{equation}
\frac{dn}{dt}=-\frac{n}{\tau}-\alpha{\cdot}n^{2}\label{eq1}
\end{equation}
where $n$ is the molecule density, $\tau$ is the molecule radiative decay lifetime and $\alpha$ is the bimolecular Penning ionization decay coefficient~\cite{Keto}. The solution for Eq.~(\ref{eq1}) is
\begin{equation}
n(t)=n_{0}\cdot\frac{e^{-t/\tau}}{1+n_{0}\alpha\tau(1-e^{-t/\tau})}\label{eq2}
\end{equation}
where $n_{0}$ is the initial molecule density. The fluorescence signal $I(t)$ as a function of time can be written as
\begin{equation}
I(t)=I_{0}\cdot\frac{e^{-t/\tau}}{1+n_{0}\alpha\tau(1-e^{-t/\tau})}\label{eq2}
\end{equation}
where $I_{0}$ is the initial fluorescence signal strength. The solid line in Fig.~\ref{molecule decay}~(b) is a least-squares fit to the measured data based on Eq.~(\ref{eq2}). The obtained molecule radiative lifetime $\tau$ is 11.6~s which is close to the previously measured one~\cite{McKinsey1999}. As one can see in Fig.~\ref{molecule decay}~(b), the decay curve bends and deviates from the straight line $e^{-t/\tau}$ decay at beginning. This is because that initially the molecule density is high and the Penning ionization decay dominates the decay process. The fitted value for quantity $n_{0}\alpha$ is 0.35~s$^{-1}$. If we take $\alpha$ to be 2.5$\times10^{-10}$~cm$^{3}$s$^{-1}$ at 2~K~\cite{Keto}, the initial molecule density $n_{0}$ is  estimated to be 1.4$\times10^{9}$~cm$^{-3}$.

\section{Proposal on visualizing the normal-fluid velocity profile in a channel}
The ability to trace the normal-fluid flow and quantitatively measure the normal-fluid velocity provides for us the chance to better
understand the role of the normal fluid in many hydrodynamical processes of superfluid helium.
For example, it has been known for many years that in a counterflow channel with small aspect ratio
there exist two different states of quantum turbulence of Helium~II denoted by Tough as T-1 and T-2
states~\cite{Tough1982}. The T-1 state appears at low values of heat flux while the T-2 state
appears at higher heat flux and is characterized by a much higher vortex line density. Melotte and
Barenghi studied the nature of the two turbulent states and proposed that this puzzle was related
to the stability of the normal fluid~\cite{Carlo1998}. In the T-1 state, the superfluid is
turbulent, but the vortex line density is not sufficient to significantly alter the laminar
Poiseuille-like profile of the normal fluid. As the heat flux is increased, eventually the line
density becomes large enough to destabilize the normal fluid and the normal-fluid flow becomes
turbulent in the T-2 state.

\begin{figure}
\begin{center}
\includegraphics[scale=0.4]{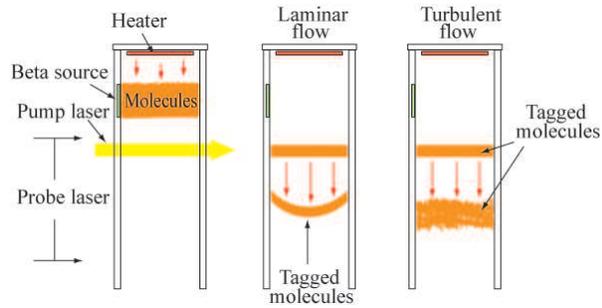}
\caption{(Color online) Schematic diagram showing the proposed experiment on imaging the
normal-fluid velocity profile in a counterflow channel using the molecule-tagging technique as
discussed in the text.}\label{channel flow}
\end{center}
\end{figure}

So far, there is no direct experimental result to confirm Melotte and
Barenghi's idea. We propose that by using the molecule tagging-imaging technique~\cite{WeiPRL}, we shall be able to
provide useful experimental information. As shown in Fig.~\ref{channel flow}, we may use a strong beta source~\cite{Rellergert} (or field emitter~\cite{WeiPRL}) to continuously produce high density He$_{2}^{*}$
molecules in a counterflow channel. A focused 910~nm pump laser pulse prepares a straight line of tagged
molecules. At an appropriate delay time, an expanded 925~nm probe laser pulse can then be used to show the
shape of the tagged-molecule line. When the normal fluid is in laminar flow, a straight molecule
line should deform to a parabolic curve due to the Poiseuille-like velocity profile of the normal
fluid in the channel. While when the normal fluid is in the turbulent state, a nearly straight
broadened molecule line should be expected due to the flat, turbulent uniform normal-fluid velocity
profile across most of the channel~\cite{Landau}. As a result, we should be able to determine the
normal-fluid velocity profile in the channel, hence distinguishing between the laminar flow and the
turbulent flow states of the normal fluid. The vortex line density needs to be measured and one
will know if the T-1 to T-2 transition of quantum turbulence is indeed coincident with the
turbulent transition of the normal fluid as suggested by Melotte and Barenghi.

\section{Conclusions}
In conclusion, we have demonstrated that metastable He$_{2}^{*}$ molecules can be used as tracers to study the dynamics of the normal fluid in superfluid $^{4}$He. The previously developed techniques for measuring the velocity of the normal fluid enable us to further explore many interesting problems such as jet impingement and the two stages of quantum turbulence in a thermal counter flow channel.
\begin{acknowledgements}
We would like to thank Prof. Joe Vinen for helpful discussions.
\end{acknowledgements}

\end{document}